\documentclass[final,5p,times,twocolumn]{elsarticle}
\usepackage{xcolor}

\usepackage{amssymb}
\usepackage{amsmath}
\usepackage{amsfonts}
\usepackage{physics}

\date{January 20, 2023}

\journal{Chaos, Solitons \& Fractals}

\begin{document}

\begin{frontmatter}

%% Title, authors and addresses

%% use the tnoteref command within \title for footnotes;
%% use the tnotetext command for theassociated footnote;
%% use the fnref command within \author or \address for footnotes;
%% use the fntext command for theassociated footnote;
%% use the corref command within \author for corresponding author footnotes;
%% use the cortext command for theassociated footnote;
%% use the ead command for the email address,
%% and the form \ead[url] for the home page:
%% \title{Title\tnoteref{label1}}
%% \tnotetext[label1]{}
%% \author{Name\corref{cor1}\fnref{label2}}
%% \ead{email address}
%% \ead[url]{home page}
%% \fntext[label2]{}
%% \cortext[cor1]{}
%% \affiliation{organization={},
%%             addressline={},
%%             city={},
%%             postcode={},
%%             state={},
%%             country={}}
%% \fntext[label3]{}

\title{Delay-induced self-oscillation excitation in the FitzHugh-Nagumo model: regular and chaotic dynamics}

%% use optional labels to link authors explicitly to addresses:
%% \author[label1,label2]{}
%% \affiliation[label1]{organization={},
%%             addressline={},
%%             city={},
%%             postcode={},
%%             state={},
%%             country={}}
%%
%% \affiliation[label2]{organization={},
%%             addressline={},
%%             city={},
%%             postcode={},
%%             state={},
%%             country={}}

\author[inst1]{Vladimir V. Semenov}
\ead{semenov.v.v.ssu@gmail.com}
\author[inst1]{Andrei V. Bukh}
\author[inst1]{Nadezhda Semenova}
\ead{semenovani@sgu.ru}

\affiliation[inst1]{organization={Institute of Physics, Saratov State University},%Department and Organization
            addressline={83 Astrakhanskaya str.}, 
            city={Saratov},
            postcode={410012}, 
            country={Russia}}

\begin{abstract}
%% Text of abstract
The stochastic FitzHugh-Nagumo model with time delayed-feedback is often studied in excitable regime to demonstrate the time-delayed control of coherence resonance. Here, we show that the impact of time-delayed feedback in the FitzHugh-Nagumo neuron is not limited by control of noise-induced oscillation regularity (coherence), but also results in excitation of the regular and chaotic self-oscillatory dynamics in the deterministic model. We demonstrate this numerically by means of simulations, linear stability analysis, the study of Lyapunov exponents and basins of attraction for both positive and negative delayed-feedback strengths. It has been established that one can implement a route to chaos in the explored model, where the intrinsic peculiarities of the Feigenbaum scenario are exhibited. For large time delay, we complement the study of temporal evolution by the interpretation of the dynamics as patterns in virtual space. 
\end{abstract}

%%Graphical abstract
%\begin{graphicalabstract}
%\includegraphics{grabs}
%\end{graphicalabstract}

%%Research highlights
%\begin{highlights}
%\item Research highlight 1
%\item Research highlight 2
%\end{highlights}

\begin{keyword}
 FitzHugh-Nagumo model \sep excitable regime \sep time delay \sep delay-induced self-oscillations \sep delay-induced chaos \sep dissipative solitons
%% PACS codes here, in the form: \PACS code \sep code
\PACS 05.45.-a \sep 05.10.-a \sep 47.54.-r \sep 02.30.Ks 
%% MSC codes here, in the form: \MSC code \sep code
\MSC[] 70K05 \sep 70K20 \sep 70K50 \sep 70L05 \sep 34K11 \sep 34K18 
%% or \MSC[2008] code \sep code (2000 is the default)
\end{keyword}

\end{frontmatter}

%% \linenumbers

%% main text
\section*{Introduction}
\label{sec:intro}
Time delay is inherently present in a wide range of systems in physics, biology, economics, and other fields \cite{ERN17}. In particular, a manifold of biological processes associated with time delay ranges from individual \cite{crook1997} and collective \cite{ORO10} neural cell activity  to the population dynamics \cite{KUA93,pal2020}. A large variety of applications associated with natural delay and imposed time-delay processes includes medical issues such as therapy for  Parkinson’s disease management \cite{rosin2011,little2013,popovych2018} and control of other pathological brain rhythms (for instance, epileptic seizures) \cite{rosenblum2004,modolo2010}, cardiac rhythm control \cite{ferreira2011}, epidemic propagation control \cite{young2019,BAU17}. In addition, time-delay processes play a significant role in chemical reaction rates \cite{schell1986,santen1972}, laser dynamics \cite{tang2001,lin2003,brunner2018,chembo2019}, analog and digital electronic circuits \cite{banerjee2018,semenov2018}, traffic flow simulation \cite{sipahi2007}, climate change modelling \cite{KEA17}.

Besides the presence of delays in biological neural networks and their influence on the neural network behaviour \cite{GAS07b,JIR08,JIR09,ORO10,WAN17}, we would like to emphasize the impact of time-delay loops in the context of artificial intelligence. Time delays are widely used for the development of delay-based machine learning algorithms involving spiking neural networks \cite{han2021,wang2019}. Long time-delay loops provide for physical realization of reservoir computing architecture \cite{tanaka2019,brunner2019}. The introduction of time-delay reservoir computing enabled simple optical, electronic and opto-electronic hardware implementations, which led to improvements of computation time scales for supervised learning \cite{martinenghi2012,brunner2018-2,larger2017,vandersande2017}. Moreover, deep-learning architectures have been adopted for the reservoir computing \cite{penkovsky2019,goldmann2020,stelzer2021}. The delay-based reservoirs were successfully applied to a wide range of tasks, such as chaotic time series forecasting or speech recognition. In addition, using a single bistable delayed-feedback dynamical node as a network, one can synthesize a physical device (for instance, see the delayed-feedback-based Ising machine \cite{boehm2019}) for solving problems with non-deterministic polynomial-time hardness (so-called NP-hard problems). 

In the context of nonlinear dynamics, time delay represents a factor being responsible for a broad spectrum of fundamental phenomena such as delay-induced regular oscillatory behavior~\cite{schoell2009time,buric2003dynamics,dahlem2009dynamics,valles2011traveling}, delay-induced bifurcations~\cite{schanz2003, janson2021} and multistability~\cite{hizanidis2008, madadi2018}, stabilization of steady states and periodic orbits~\cite{PYR92, hoevel2005}, vibrational resonance~\cite{YAN10b, HU12}, delay-induced chaos~\cite{farmer1982,krupa2016}, delayed-feedback control of chaos~\cite{PYR92} (including spatiotemporal patterns~\cite{beck2002, unkelbach2003}), delay-induced synchronization~\cite{marti2003, klinshov2013, nathe2020} and desynchronization~\cite{popovych2005, popovych2017}. In addition to the deterministic effects, time delay represents a useful approach for controlling stochastic phenomena such as stochastic \cite{WU08a,mei2009,jia2009} and coherence \cite{JAN04,BAL04,PRA07,AUS09,BRA09,geffert2014,SEM15} resonances, noise-induced chimera states in networks of nonlocally-coupled excitable oscillators \cite{ZAK17a} and noise-induced patterns in excitable media \cite{balanov2006}. Thus, delayed feedback is a powerful tool for achieving a wide range of operating regimes, enhancing amplitude-frequency characteristics, and controlling the dynamics of nonlinear systems \cite{schoell2008,sun2013, flunkert2013}.

As was first mentioned in Ref. \cite{arecchi1992}, there exists an analogy between the behavior of time-delayed systems and the dynamics of ensembles of coupled oscillators or spatially extended systems \cite{giacomelli1996}. The similarity takes place when the delay time, $\tau$, is much longer than system's response time, which allows the system to reveal spatio-temporal phenomena (for example, coarsening \cite{giacomelli2012}, chimera states \cite{semenov2016,larger2015,larger2013}, soliton dynamics \cite{brunner2018,semenov2018,yanchuk2019}) in the purely temporal dynamics of a time-delay system. This space-time analogy can be obtained by implementing a space-time transformation of the delay-feedback system, where the temporal dynamics is mapped onto space-time ($\sigma$, $n$) by introducing the space-time map $t=n\tau+\sigma$ with an integer (slow) time variable $n$, and a pseudo-space variable $\sigma \in [0,\eta]$, where $\eta=\tau+\delta$ with a small quantity $\delta$. For each set of parameters a unique value $\eta$ can be chosen such that the oscillatory dynamics is periodic with the period $\eta$. The mentioned technique is described in details in review \cite{yanchuk2017} as well as a manifold of spatio-temporal phenomena tracked down in the dynamics of delayed-feedback oscillators. In addition, the large delay time in one delay-feedback system can be interpreted as one-directional impact of the identical system on the considered oscillator without any feedback \cite{ZAK17a, njitacke2022extremely}. In this case one may expect appearance of self-sustained oscillations. 

It is known that delayed feedback provides for the realization of self-oscillatory dynamics being unachievable in the absence of delay. In particular, it has been established that delayed feedback loops allow to shift critical parameter values corresponding to the subcritical Andronov-Hopf bifurcation. This fact became a base for the delayed-feedback-based coherence resonance control scheme in non-excitable systems \cite{geffert2014,SEM15}. In contrast, the mechanism of noise-induced oscillation control by using delayed-feedback in the FitzHugh-Nagumo system (which is an excitable oscillator) in the regime of coherence resonance is not associated with changing the critical parameter value and the occurrence of the supercritical Andronov-Hopf bifurcation \cite{PRA07,JAN04}. The delay-induced self-oscillation excitation in the FitzHugh-Nagumo neuron has been studied in the context of deterministic dynamics for the short-time-delay limit \cite{ERN16} as well as for noticeable time delay \cite{weicker2016,romeira2016}. In all the cases the time delay feedback induces bistability as the coexistence of the regular self-oscillatory dynamics (spiking) and a quiescent steady state regime, whereas the stable equilibrium is a single attractor at zero delay.

In the current study, we summarize the materials published in Refs. \cite{ERN16,weicker2016,romeira2016} and complement them by our results which describe the FitzHugh-Nagumo oscillator with time-delayed feedback for both negative and positive delayed-feedback strength. Based on the linear stability analysis of steady states and numerical modelling of the equations under study, we demonstrate that the action of delayed-feedback loop can result in oscillations arising in the FitzHugh-Nagumo model accompanied with loss of stability of equilibrium point. In addition, we analyse basins of attractions, the probabilities to observe the self-oscillations, and the evolution of Lyapunov exponents. Besides the delay being smaller or comparable with the system response time, we also consider the case of long delay. By this way, we combine the study in the context of temporal evolution with the research of structures in quasi space.

%In the framework of the time-delayed control, the FHN system is often studied in excitable regime in the presence of noise and time-delayed feedback. In such a case the last one may control the noise induced oscillations. Here we show that the time-delayed feedback not only controls the existing oscillations, but also may induce new regimes and shift the bifurcation values of control parameters. 

%In this paper we summarize the existing and new results on the regimes arising in excitable FHN systems with time-delayed feedback. We confirm the results obtained in numerical simulation by means of linear stability analysis, probability analysis, calculation of basins of attraction, and Lyapunov exponents. The effect of time-delay is studied for both positive feedback and negative feedback.

\section{System under study}\label{sec:system}
The FitzHugh-Nagumo system originally proposed for the description of processes in nerve membranes \cite{fitzhugh1961,nagumo1962} is a paradigmatic model for the type-II excitability. In the present paper, we consider the FitzHugh-Nagumo model with time-delayed feedback: 
\begin{equation}\label{eq:system} 
\begin{array}{l} 
\varepsilon\dot{x} = x-x^3/3-y + \gamma(x_\tau - x), \\ 
\ \dot{y}=x+a, 
\end{array}
\end{equation} 
where $x=x(t)$ is the activator variable, $y=y(t)$ is the inhibitor variable. A parameter $\varepsilon\ll 1$ is responsible for the time scale separation of fast activator and slow inhibitor variable, $a$ is the threshold parameter which determines the system dynamics: the system exhibits the excitable regime at $|a|>1$ and the oscillatory one for $|a|<1$. A parameter $\tau$ is the delay time, $x_{\tau}$ is a value of the variable $x$ at the time moment ($t-\tau$). A parameter $\gamma$ is the feedback strength. The feedback term $\gamma(x_\tau - x)$ was first introduced by K. Pyragas to stabilize unstable periodic orbits resulting in the stabilization of the chaotic dynamics \cite{PYR92}, but is also  used for controlling various kinds of the deterministic and stochastic dynamics (see the references in the introduction). In this paper, we consider the FitzHugh-Nagumo system in the excitable regime ($a>1$ and $\varepsilon>0$) for varying delay time and time-delayed feedback strength. 

Besides the analytical approaches described in appendices, our results are based on numerical simulations carried out by the integration of the model equation by using the Heun method \cite{mannella2002} with time step $h=0.001$ starting from random or specially prepared initial conditions ($x_0$,$y_0$), where $x_0=x(t\in[-\tau:0])=\mathrm{const}$.

Using the term 'initial conditions' ($x_0$,$y_0$) we mean the following. When integrating Eqs. (\ref{eq:system}) during the numerical modelling, a value of variable $x$ at the time moment $(t+h)$ is determined by the previous values of the dynamical variables, $x(t)$ and $y(t)$, as well as by a value $x(t-\tau)$ (here, $h$ is the integration time step). This means that the integration of system (\ref{eq:system}) starting from the initial time moment $t=0$ requires the initialization of $\tau/h$ values of  $x$ in the time range $[-\tau:0)$. We set all of them to be identical and equal to $x_0$. In addition, we initialize a single initial value of the $y$-variable, $y_0=y(t=0)$ for starting the numerical simulations.

%In this article we mainly use the values of the parameters $a>1$ and $\varepsilon>0$ corresponding to excitable regime. 

%The initial conditions which will be further used are set in the next way. The initial condition $(x_0,y_0)$ corresponds to the case when $(x,y)=(x_0,y_0)$ for all $-\tau\le t<0$.

%\textcolor{red}{+ TIME DELAY $\to$ TO SPACE}

\section{Positive time-delayed feedback strength ($\gamma>0$)}
\label{sec:TDF_positive}
\subsection{Temporal dynamics}
%\begin{itemize}
%    \item The probability that there is anything besides a state of equilibrium.
%    \item $0<P<1$ $\to$ multistability
%    \item basins of attraction + main regimes
%    \item Are these oscillations periodic or chaotic? $\to$ Lyapunov exponents
%    \item What if we have large time delay? Solitons.
%\end{itemize}

\begin{figure}[t] 
\centering{\includegraphics[width=1\linewidth]{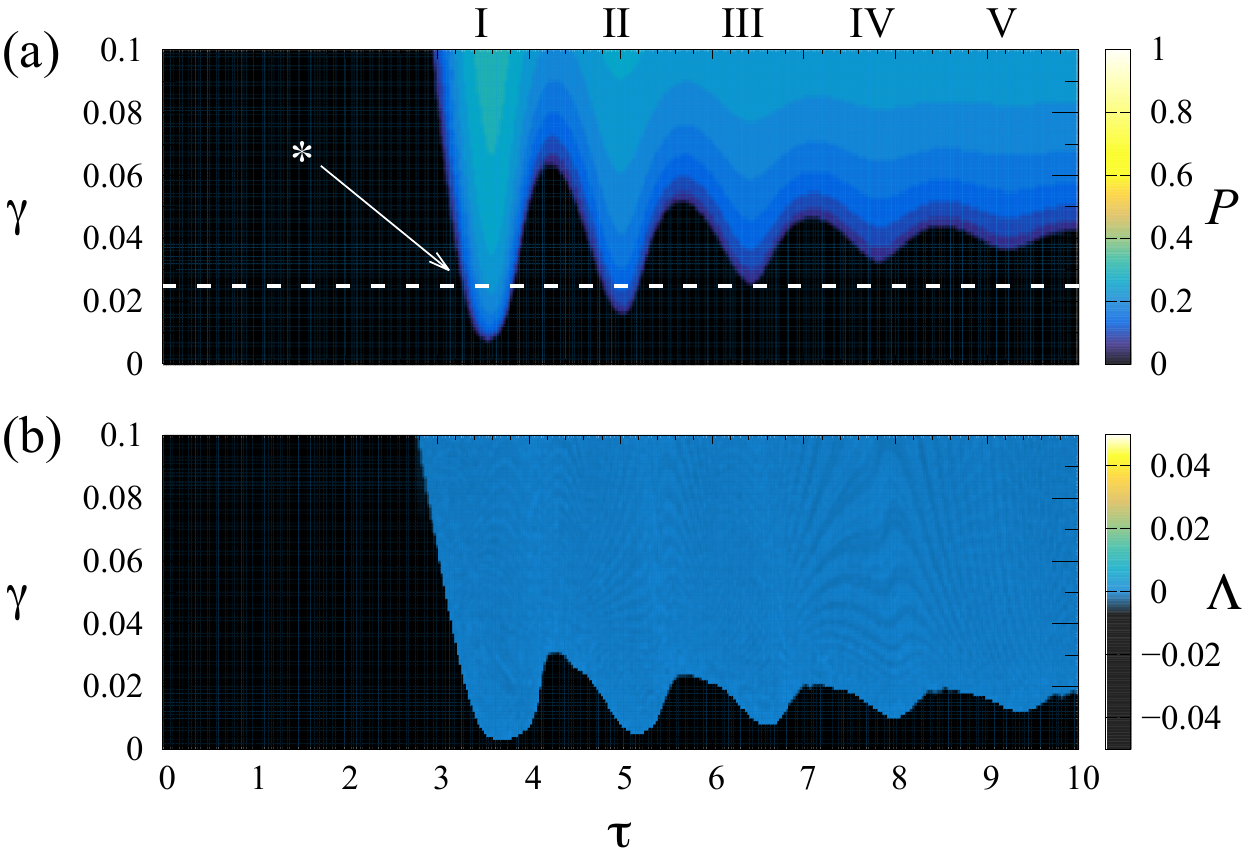}} 
\caption{Probability $P$ of detecting regimes being different from the equilibrium ($x_s,y_s$) in model (\ref{eq:system}) and maximal Lyapunov exponent $\Lambda$ in ($\tau$,$\gamma$)-plane. Parameters: $\varepsilon=0.05$, $a=1.001$.
%(a) Probability of the occurrence of the regimes different from the equilibrium ($x_s,y_s$) in system (\ref{eq:system}) on the parameter plane ($\gamma$,$\tau$); (b) Lyapunov exponent in the same parameter plane. Parameters are: $\varepsilon=0.05$, $a=1.001$.
}\label{fig:g_pos_prob_lyap} 
\end{figure}

As noted above, we consider system (\ref{eq:system}) in the excitable regime, occurring when the threshold parameter $a$ is greater than unity. In this case the system (\ref{eq:system}) without time-delayed feedback is characterised by the presence of a single attractor in the phase space: a stable focus with coordinates $x_s=-a$, $y_s=\frac{a^3}{3}-a$. However, in the presence of time delay, one can observe trajectories tracing self-oscillations which are not associated with the initial single attractor. The possibility to realize the self-oscillatory dynamics depends both on the delayed-feedback parameters ($\gamma$ and $\tau$) and the initial conditions. In order to quantitatively characterize this fact, we introduce the probability $P$ to achieve the regimes where the phase point oscillates outside the vicinity of the equilibrium point $(x_s,y_s)$. 
The corresponding probability is given in Fig.~\ref{fig:g_pos_prob_lyap}(a) for varying feedback strength $\gamma$ and delay time $\tau$ and fixed parameters $a=1.001$ and $\varepsilon=0.05$. Here, the probability for each $\gamma$ and $\tau$ is calculated by modelling the system 1500 times starting from the grid initial conditions $x_0\in[-2.5;2.5]$ and $y_0\in[-1.5;1.5]$, where $x(t\in[-\tau:0])=x_0=\mathrm{const}$. To eliminate the action of transients, the integration time is sufficiently long, $t_{\text{total}}=10000$. After each integration cycle, the position of the phase point was analyzed for proximity to the equilibrium $(x_s,y_s)$. Finally, a number of trajectories outside the vicinity of equilibrium was divided by the total number of initial conditions. This method for calculating probability is called 'phase space probability' in some literature, where a set of virtual copies of considered system starting from different initial conditions is called a statistical ensemble, and averages computed with this definition of probability are called ensemble averages \cite{Eastman2014}.

Figure \ref{fig:g_pos_prob_lyap}(a) shows a large area corresponding to $P=0$ and coloured in black. In this area, we observe the only quiescent steady state regime associated with the equilibrium $(x_s,y_s)$. Outside this area the probability takes values in the range  $P\in(0:1)$, which indicates the coexistence of the steady state regime with undamped oscillations. 

\begin{figure}[t] 
\centering{\includegraphics[width=1\linewidth]{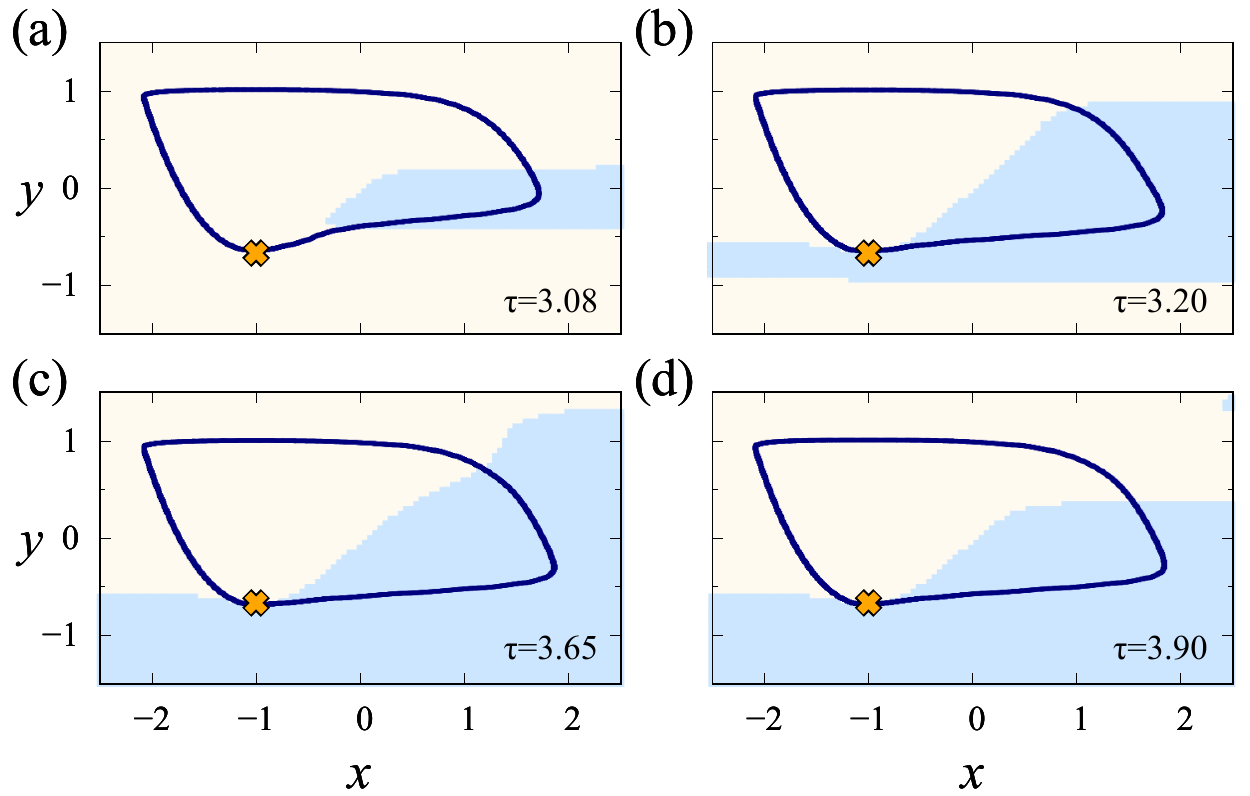}} 
\caption{Projections of two attractors on ($x,y$) phase plane obtained by increasing delay time $\tau$ through the boundary between black and blue areas in Fig.~\ref{fig:g_pos_prob_lyap}(a, $\ast$-point). Light-blue and yellow areas depict the section of the basins of attraction in the functional space for blue limit cycle and orange equilibrium point, respectively. Parameters: $\gamma=0.025$, $\varepsilon=0.05$, $a=1.001$.
% Main regimes and their basins of attraction on the boundary between black and blue areas in Fig.~\ref{fig:g_pos_prob_lyap}(a) ($\ast$-point) for the delay time $\tau=3.08 $ (a), $\tau=3.2$ (b), $\tau=3.65$ (c), $\tau=3.9$ (d). System's (\ref{eq:system}) parameters are: $\gamma=0.025$, $\varepsilon=0.05$, $a=1.001$. Light-blue areas correspond to the basins of attraction for the limit cycles denoted by blue curves, while yellow regions complies the basins of attraction for the equilibrium point (orange x-point).
}\label{fig:g_pos_basins} 
\end{figure}

The multistability of a system without time-delayed feedback can be visualized by attractors in a phase space and their basins of attraction, showing a set of initial conditions leading to corresponding attractor. The system under consideration is of infinite dimension, and therefore we can show only a section of the basins of attraction in the functional space and projection of phase trajectories on the phase plane ($x,y$). Figure \ref{fig:g_pos_basins} shows these results for different $\tau$ values. The section of the basins of attraction was obtained in the next way. The selected ranges of ($x,y$) values were divided into a lattice of initial conditions ($x_0,y_0$). It is described in the last paragraph of Sect.~\ref{sec:system} how the initial conditions are specified. The initial conditions leading to stable steady state shown by orange cross are colored in yellow, while the initial conditions leading to the blue attracting set are colored in light blue.

In particular, Fig. \ref{fig:g_pos_basins} illustrates the evolution of attracting sets by projections on phase plane ($x,y$) and corresponing section of basins of attraction for fixed feedback strength, $\gamma=0.025$ (this level is indicated by the dashed line in Fig.~\ref{fig:g_pos_prob_lyap} (a)) and increasing time delay starting from the value $\tau=3.08$ (marked by the symbol '$\ast$' in Fig.~\ref{fig:g_pos_prob_lyap} (a)). The coexistence of two attractors is illustrated in Fig. \ref{fig:g_pos_basins}. The first attractor, the stable steady state (the orange cross) is characterised by yellow basin of attraction. The second attractor, a stable limit cycle (the blue curve), is characterised by the basin coloured in light-blue. As demonstrated in Fig. \ref{fig:g_pos_basins}, the section of basins of attraction evolve with increasing time delay. 

The maximal size of the limit cycle's basin can be found in the center of area I in Fig.~\ref{fig:g_pos_prob_lyap}(a). The corresponding section of basin of attraction is shown in Fig.~\ref{fig:g_pos_basins}(c). The further increase of the parameter $\tau$ inside part I leads to decrease of the basin (Fig.~\ref{fig:g_pos_basins}(d)) and its total disappearance when $\tau$ comes back again to the black area of Fig.~\ref{fig:g_pos_prob_lyap}(a). The same birth of oscillations and their disappearance can be observed in the next four parts: II, III, IV, V.

\begin{figure}[t] 
\centering{\includegraphics[width=1\linewidth]{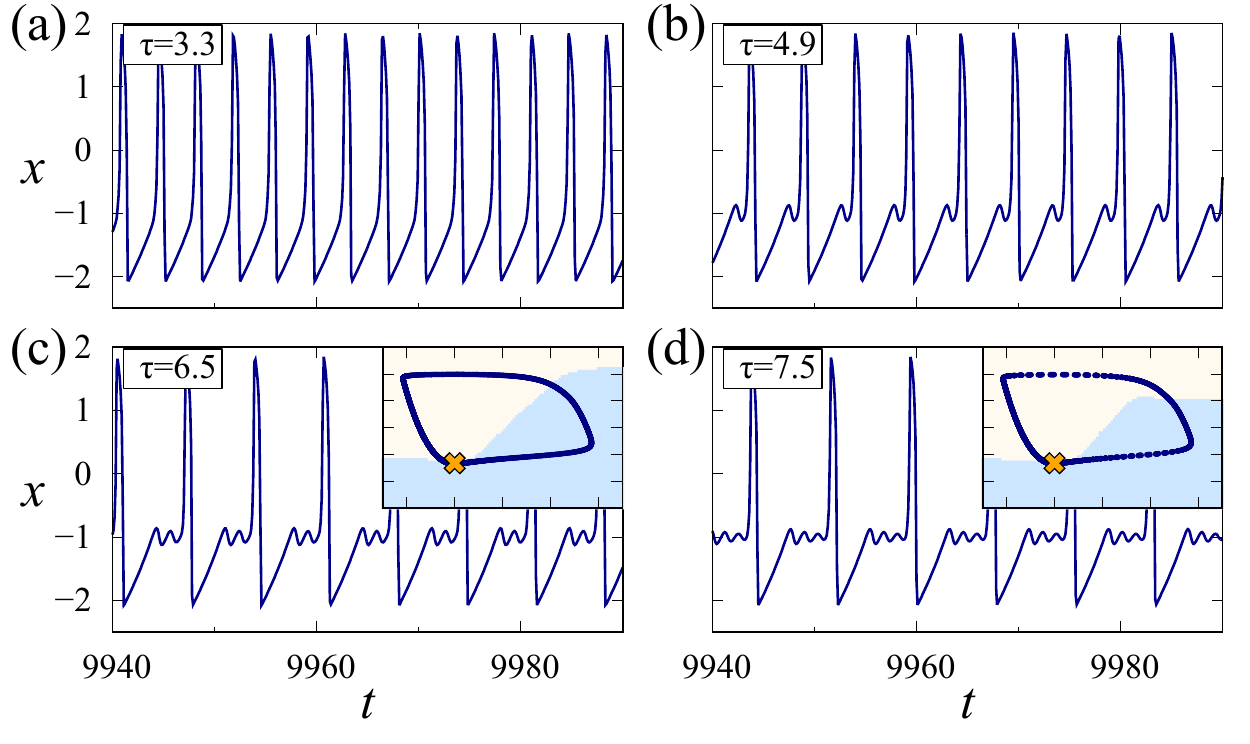}} 
\caption{Temporal dynamics of model (\ref{eq:system}) with different period numbers. The regimes shown in panels (c,d) correspond to coexistence of the stable focus point and the limit cycle indicated in the same manner as in Fig.~\ref{fig:g_pos_basins}. Parameters: $\gamma=0.03$, $\varepsilon=0.05$, $a=1.001$.
% Temporal dynamics of system (\ref{eq:system}) with different period numbers. The inserted phase partraits demonstrate two coexisting attractors of stable focus point (orange x-point) and limit cycle (blue curve) with corresponding basins of attractions shown by similar light colors. Parameters: $\tau=3.3$ (a), $\tau=4.9$ (b), $\tau=6.5$ (c), $\tau=7.5$ (d). Other parameters: $\gamma=0.03$, $\varepsilon=0.05$, $a=1.001$.
}\label{fig:g_pos_cycles} 
\end{figure}

In addition to the visual distinction between areas I--V in Fig.~\ref{fig:g_pos_prob_lyap}(a), they differ in the type of oscillations. The spiking behaviour with period-1 limit cycle can be observed in the part I of blue area in Fig.~\ref{fig:g_pos_prob_lyap}(a). The corresponding time realization is given in Fig.~\ref{fig:g_pos_cycles}(a). With further increasing the delay time $\tau$ and moving further into the parts II, III, IV, the number of periods is growing as period-2 oscillations (Fig.~\ref{fig:g_pos_cycles}(b)) in Part II, period-3 oscillations (c) in Part III, and  period-4 oscillations (Fig.~\ref{fig:g_pos_cycles}(d)) in Part IV. A similar pattern persists with a further increase in the parameter $\gamma$.

The existence of observed oscillatory regimes is confirmed by calculation of maximal Lyapunov exponents (Fig.~\ref{fig:g_pos_prob_lyap}(b)) on the same parameter plane ($\tau$,$\gamma$). The map of Lyapunov exponents is prepared in the next way. The main Lyapunov exponent is determined as follows

\begin{equation}\label{eq:varsystem} 
\Lambda = \overline{\lim_{T\to\infty}} \dfrac 1T \ln \dfrac{||\boldsymbol{\xi}(t=T)||}{||\boldsymbol{\xi}(t=0)||},
\end{equation}
where the initial length $||\boldsymbol{\xi}(t=0)||$ is set to be equal to unity, while the final length $||\boldsymbol{\xi}(t=T)||$ is obtained by simulating the system (\ref{eq:system}) together with the following system of equations:
\begin{equation}\label{eq:system2} 
\left\{ \begin{array}{l} 
\varepsilon\dot{\xi_1} = \xi_1 - x^2\xi_1 - \xi_2 + \gamma({\xi_1}_\tau - \xi_1), \\ 
\ \dot{\xi_2}=\xi_1, 
\end{array}\right. 
\end{equation}
where $\xi_1,\xi_2$ are the components of the vector $\boldsymbol{\xi}$, and parameters $\varepsilon, \gamma, \tau, x, y$ have the same meaning as in Eq.~(\ref{eq:system}). 

The black areas in Fig.~\ref{fig:g_pos_prob_lyap}(b) with negative maximal Lyapunov exponents $\Lambda$ correspond to the existence of only stable equilibrium point without any oscillations. The blue region with $\Lambda=0$ confirms the existence of periodic oscillations for chosen parameters $\gamma$ and $\tau$.

We have calculated the Lyapunov exponents in a wider range of the feedback strength $\gamma\in[0;1]$ and have found that the previously obtained trend remains. Only two regimes were observed in the considered parameter plane including stable equilibrium point with $\Lambda<0$ for small $\gamma$ and $\tau$ and its coexistence with periodic oscillations with $\Lambda=0$ for the rest of the parameter values. Regimes with positive $\Lambda$ have not been found for $\gamma>0$, which confirms that the delay-induced self-oscillations are regular. These calculations are given in \ref{sec:appendix:lyaps_gamma_positive}.

\subsection{Soliton structures at large time delay}\label{sec:solitons}

\begin{figure}[t] 
\centering{\includegraphics[width=0.9\linewidth]{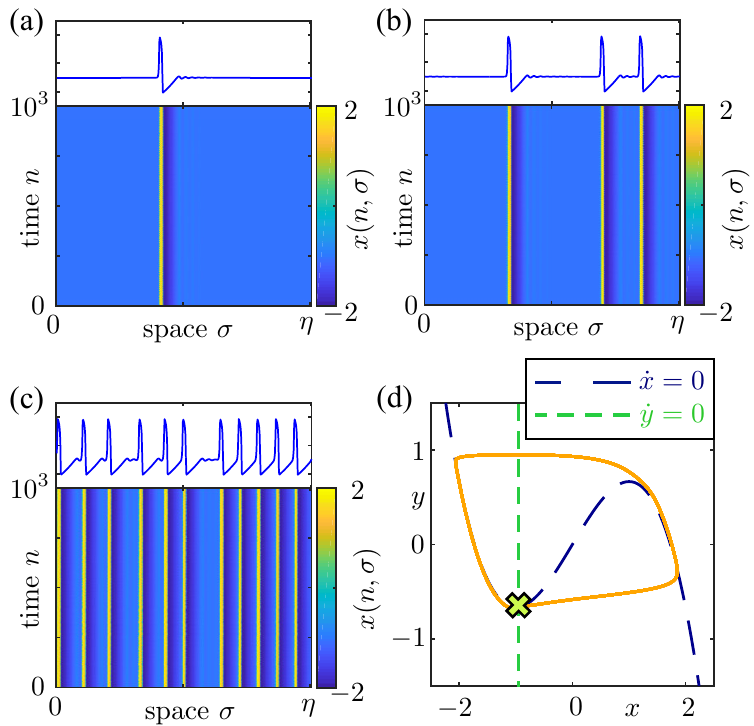}} 
\caption{Single-soliton (panel (a)) and multiple-soliton (panels (b) and (c)) regimes revealed in the temporal dynamics of model (\ref{eq:system}) at positive delay feedback strength. The upper insets in panels (a)-(c) show the system state in quasi-space $\sigma$ at the last discrete time moment $n=10^3$. Panel (d) show the projection of the multiple-soliton pattern attractor from panel (c) on ($x,y$) phase plane. Parameters: $\gamma=0.05$, $\varepsilon=0.05$, $a=1.01$, $\tau=50$, $\eta=50.211$ (panels (a,b)), $\eta=50.2197$ (panel (c)).
% Single-soliton (panel (a)) and multiple-soliton (panels (b) and (c)) regimes revealed in the temporal dynamics of model (\ref{eq:system}) at positive delay feedback strength. The upper insets in panels (a)-(c) show the system state in quasi-space $\sigma$ at the last discrete time moment $n=10^3$. Phase portrait in panel (d) corresponds to the multiple-soliton pattern in panel (c). Parameters are: $\gamma=0.05$, $\varepsilon=0.05$, $a=1.01$, $\tau=50$, $\eta=50.211$ (panels (a,b)), $\eta=50.2197$ (panel (c)).
}\label{fig:long_delay_positive} 
\end{figure}

If the delay time is sufficiently long, one can complement studying model (\ref{eq:system}) in the frameworks of the temporal evolution by the consideration of the system in the context of spatio-temporal structures. In particular, the mentioned approach applied to Eqs. (\ref{eq:system}) with positive delay feedback strength allows to reveal soliton structures \cite{romeira2016}. It is important to note that such structures are observed in the nonlinear dissipative system, can persist in the presence of noise, save their shape and do not decay over time. For these reasons, they can be referred to the dissipative solitons. Choosing the initial conditions or using a temporal external influence, one can induce single [Fig. \ref{fig:long_delay_positive}(a)] and multiple [Fig. \ref{fig:long_delay_positive}(b)-(c)] dissipative solitons in system (\ref{eq:system}). The oscillatory regimes illustrated in Fig. \ref{fig:long_delay_positive}(a)-(c) represent a particular kind of delay-induced self-oscillatory dynamics: the observed oscillations are obtained at $a=1.01$ where the self-oscillations cannot be exhibited in the absence of time delay. On projections on the phase plane, the soliton structures in Fig. \ref{fig:long_delay_positive}(a)-(c) consist of self-sustained alternating spike loops and the rest states at the equilibrium [Fig. \ref{fig:long_delay_positive}(d)]. 

The multiplicity of the dissipative solitons can be potentially used in the context of information storage. Indeed, adjusting the initial conditions, any number of solitons can be disposed in the virtual space and vice versa, each assigned soliton can be canceled by a slight external impact. However the minimal distance between the units is restricted. This is a limitation which bounds the maximal possible number of solitons in space. Nevertheless, the mentioned restriction can be overcame by increasing time delay. Then the information capacity of the system can grow.

\section{Negative time-delayed feedback strength ($\gamma<0$)}
\label{sec:TDF_negative}
%\begin{itemize}
%    \item The probability that there is anything but a state of equilibrium.
%    \item P is 0 or 1. Therefore, there is no multistability and we can make linear stability analysis
%    \item main regimes. Periodic oscillations and different number of loops.
%    \item There is something like chaotic behaviour $\to$ Lyapunov exponents/
%    \item Chaotic realizations and phase portraits.
%    \item What if we have large time delay? Spatial chaos?
%\end{itemize}

\begin{figure}[t] 
\centering{\includegraphics[width=1\linewidth]{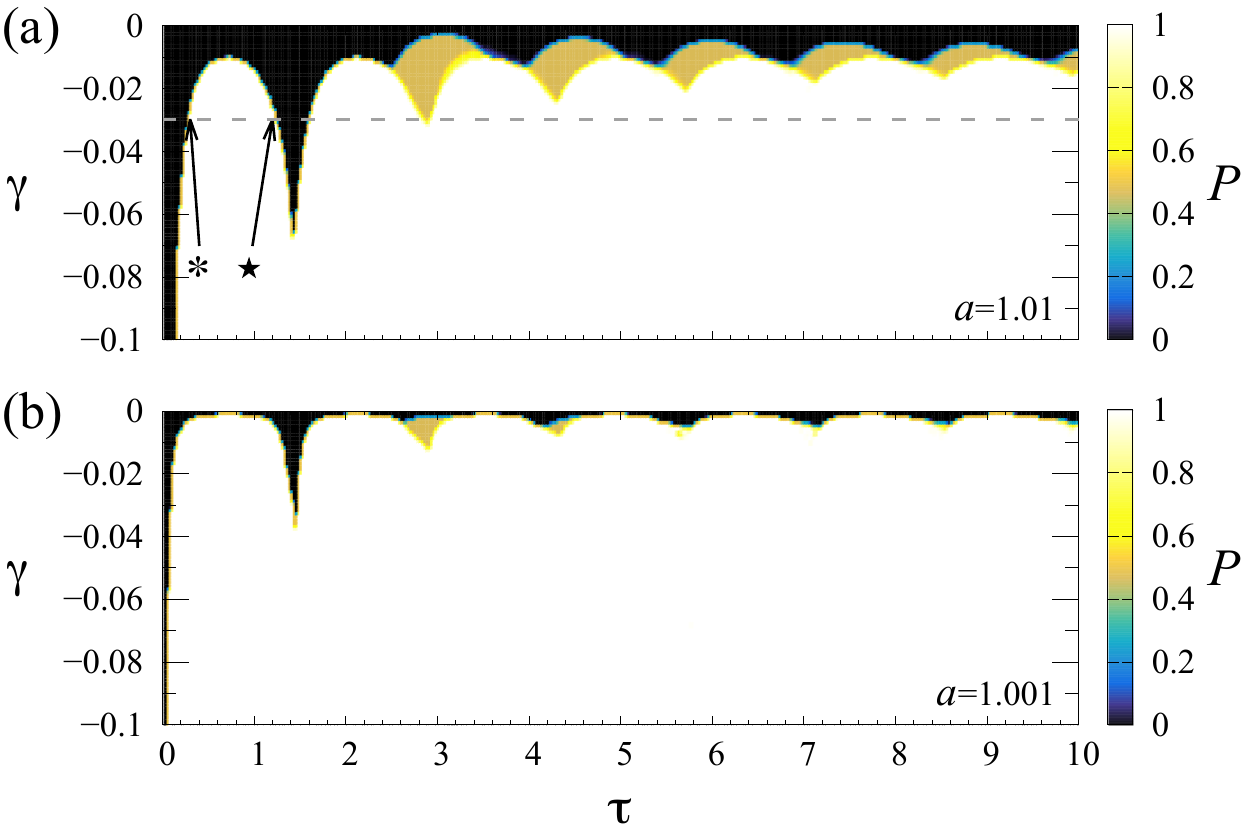}} 
\caption{Probability $P$ of detecting regimes being different from the equilibrium ($x_s,y_s$) in model (\ref{eq:system}) in ($\tau$,$\gamma$)-plane. Black area: only stable equilibrium; white region: limit cycles. Parameters: $\varepsilon=0.05$, $a=1.01$ (a) and $a=1.001$ (b).
% Parameter plane ($\tau$, $\gamma$) showing the probability of finding the regimes different from the equilibrium ($x_s,y_s$) in system (\ref{eq:system}) at $\varepsilon=0.05$, $a=1.01$ (panel (a)) and $a=1.001$ (panel (b)).
}\label{fig:g_neg_prob} 
\end{figure}

Let us consider the regimes induced by the time-delayed feedback with negative strength $\gamma$. As the first step, we use the same technique as in Sec.~\ref{sec:TDF_positive} to find out any exit of trajectory outside the vicinity of the equilibrium point ($x_s$,$y_s$) using the probability analysis. The corresponding probability on the plane ($\gamma$,$\tau$) is shown in Fig.~\ref{fig:g_neg_prob} for two parameter values $a=1.001$ panel (a)) and $a=1.01$ (panel (b)). There is the same color scheme as in Sec.~\ref{sec:TDF_positive}, but comparing Figs.~\ref{fig:g_pos_prob_lyap}(a) and \ref{fig:g_neg_prob}, one can see a clearly pronounced difference in maximum probability values. In contrast to probabilities calculated for $\gamma>0$ in Sect.~\ref{sec:TDF_positive}, now for $\gamma<0$, there are clear parameter areas with $P\to 1$ and $P\to 0$. It means, that there is no bistability for $\gamma<0$. In the black area of parameter values, the system demonstrates a stable equilibrium point while the parameters from the white area lead to a completely different dynamics. The probability was prepared again with global approach with listing possible initial conditions $x_0\in[-2.5;2.5]$ and $y_0\in[-1.5;1.5]$, and, therefore, the probability $P=1$ means that no trajectories come to the vicinity of the equilibrium point.

\begin{figure}[t] 
\centering{\includegraphics[width=1\linewidth]{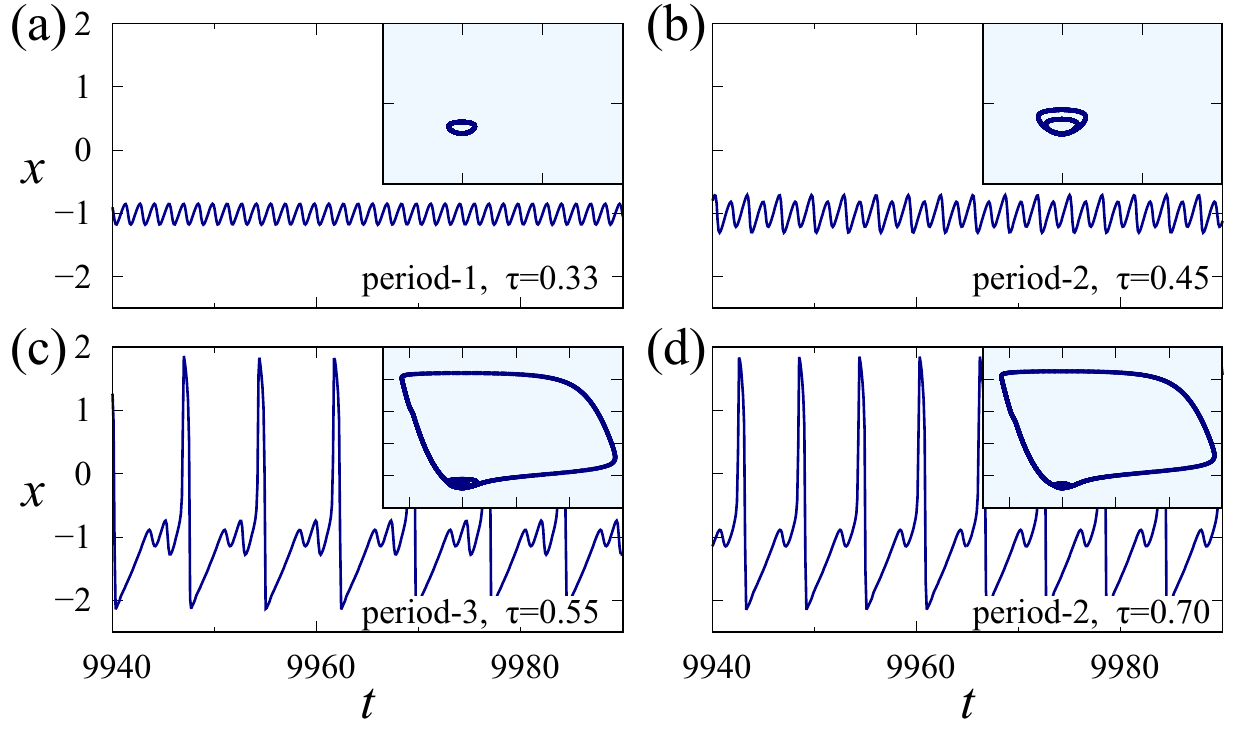}} 
\caption{Temporal dynamics of model (\ref{eq:system}) with different period numbers induced by the feedback with negative strength $\gamma$. Subpanels contain the projections of corresponding attractors on ($x,y$) phase plane. Parameters: $\gamma=-0.03$, $\varepsilon=0.05$, $a=1.01$.
% Time realizations $x(t)$ and phase portraits displaying main types of delay-induced oscillations near the equilibrium point with period-1 limit cycle (panel (a)) for $\tau=0.35$ and period-2 cycle (panel (b)) for $\tau=0.45$; spiking behavior with period-3 cycle (panel (c)) for $\tau=0.55$ and period-2 cycle (panel (d)) for $\tau=0.7$. Other parameters: $\varepsilon=0.05$, $\gamma=-0.03$, $a=1.01$.
}\label{fig:g_neg_realizations} 
\end{figure}

In order to uncover the mechanism of birth of oscillations caused by the feedback with negative strength we fix the parameter $\gamma=-0.03$ (dashed line in Fig.~\ref{fig:g_neg_prob}(b)) and move from left to right increasing the delay time $\tau$. First, in the black area for $\tau\approx 0$ only the stable equilibrium point is observed. Next, in the beginning of the white area marked by the symbol $\ast$ in Fig.~\ref{fig:g_neg_prob}(b), a limit cycle of period-1 with a small amplitude is born around the previously stable equilibrium point. The corresponding temporal realization with corresponding attracting set in projection on phase plane ($x,y$) is given in Fig.~\ref{fig:g_neg_realizations}(a). The amplitude of the obtained limit cycle grows with distancing from purple area. Therefore, the appearance of small yellow areas in Fig.~\ref{fig:g_neg_prob} is caused not by the multistability, but by the smallness of the limit cycle which is comparable with the size of the vicinity used for probability calculations. These oscillations have large decay, and finally end with equilibrium point.

At some point, the period doubling bifurcation takes place, and a new loop of the limit cycle appears (see Fig.~\ref{fig:g_neg_realizations}(b) for $\tau=0.45$). In the center of the white area situated between symbols $\ast$ and $\star$ the period-3 cycle can be obtained (see Fig.~\ref{fig:g_neg_realizations}(c) for $\tau=0.55$). This regime combines oscillations near the previous fixed point and the spiking behaviour. With increasing $\tau$ and getting close to $\star$ from the left, the number of loops decreases and period-2 can be obtained (Fig.~\ref{fig:g_neg_realizations}(d), $\tau=0.7$) with final disappearance of any oscillations when $\tau$ gets to the black area again.

With further increase of $\tau>1.5$ the system demonstrates a wide range of regimes: cycles with different period numbers and spiking behavior, like in Fig.~\ref{fig:g_neg_realizations}(a--d). The spiking behavior is the most typical regime in the system with negative time-delayed feedback. 

Such loss of stability of the equilibrium point is proven by the linear stability analysis (see \ref{sec:appendix:LSA_gamma_negative}). 

\begin{figure}[b!] 
\centering{\includegraphics[width=1\linewidth]{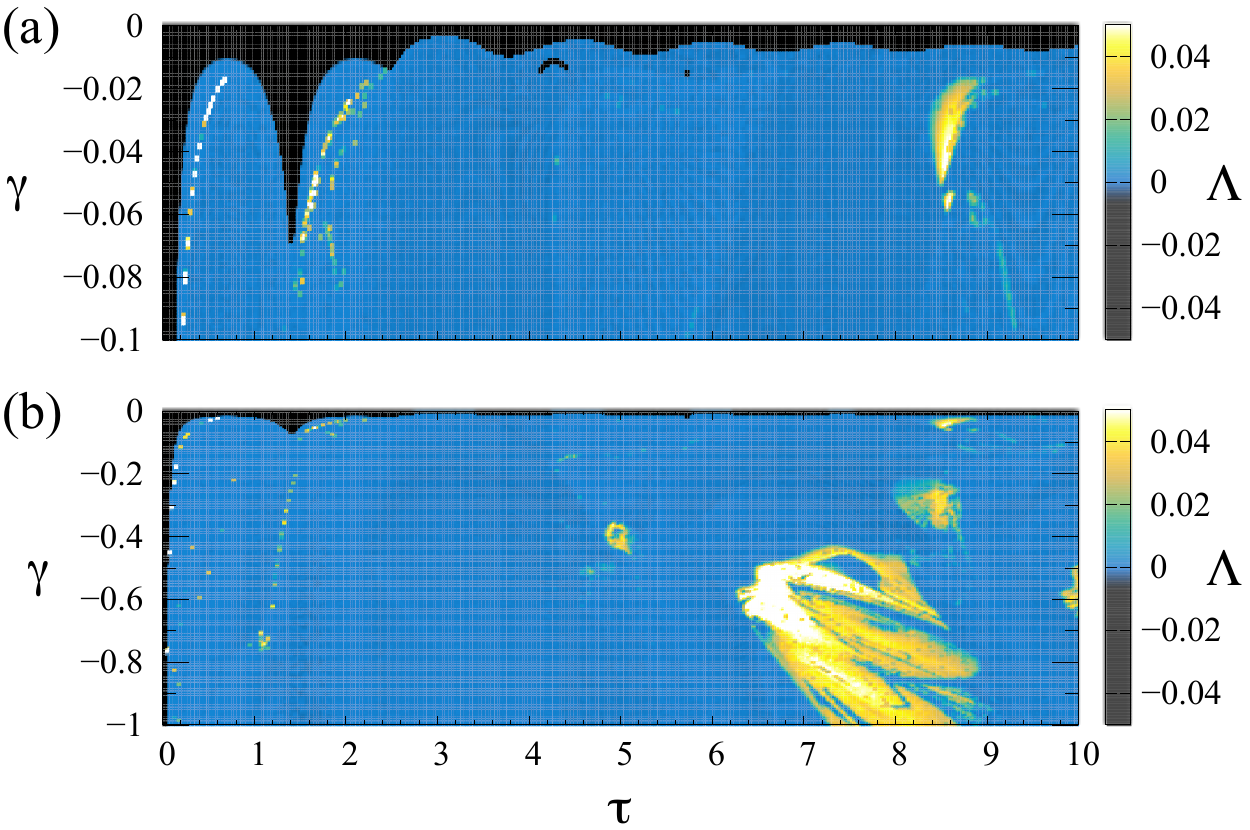}} 
\caption{Maximal Lyapunov exponent $\Lambda=\Lambda_1$ in ($\tau$,$\gamma$)-plane calculated for model (\ref{eq:system}) with negative feedback strength for two different ranges of $\gamma$ values. Black area: stable equilibrium point; blue region: limit cycles; yellow or white areas: chaotic regimes. Parameters: $\varepsilon=0.05$, $a=1.01$.
% Parameter plane ($\tau$, $\gamma$) showing the values of the maximal Lyapunov exponent $\Lambda=\Lambda_1$ in system (\ref{eq:system}) in case of negative delay feedback strength. Black area corresponds to the stable equilibrium, yellow (light-grey) area confirms to limit cycles, and yellow or white areas contain chaotic regimes. Parameters: $\varepsilon=0.05$, $a=1.01$ (a), and $a=1.001$ (b).
}\label{fig:g_neg_lyap} 
\end{figure}

The obtained loss of stability has also been proven by the Lyapunov exponents (Fig.~\ref{fig:g_neg_lyap}) described first in Sect.~\ref{sec:TDF_positive}. The existence of the stable fixed point in the black area is characterised by $\Lambda<0$, while the rest part (blue) of parameter plane $\tau\in[0;10]$, $\gamma\in [-0.1;0]$ has $\Lambda=0$, corresponding to periodic oscillations. At the same time, if we enlarge the feedback strength in absolute value, there are additional interesting regimes accompanied by $\Lambda>0$, which indicated the chaotic behaviour (yellow regions in Fig.~\ref{fig:g_neg_lyap}).

\subsection{Large feedback strength. Chaotic oscillations}

The largest part of Fig.~\ref{fig:g_neg_lyap}(a) is colored in blue corresponding to periodic oscillations with $\Lambda\approx 0$. The main difference of Lyapunov exponents calculated for positive and negative $\gamma$ (Fig.~\ref{fig:g_pos_prob_lyap}(a) and Fig.~\ref{fig:g_neg_lyap}) is that the parameter plane contains yellow regions with $\Lambda>0$ for negative $\gamma$. This indicates that chaotic oscillations can be observed for corresponding parameters $\gamma$ and $\tau$ from that regions. At the selected scale of parameter $\gamma$ (Fig.~\ref{fig:g_neg_lyap}(a)), this covers a very small parameter area. However, when the absolute value of parameter $\gamma$ is increased (Fig.~\ref{fig:g_neg_lyap}(b)), the number of such areas and their size grows. In this subsection, we take a closer look at these chaotic regimes.

\begin{figure}[t]
\centering{\includegraphics[width=1\linewidth]{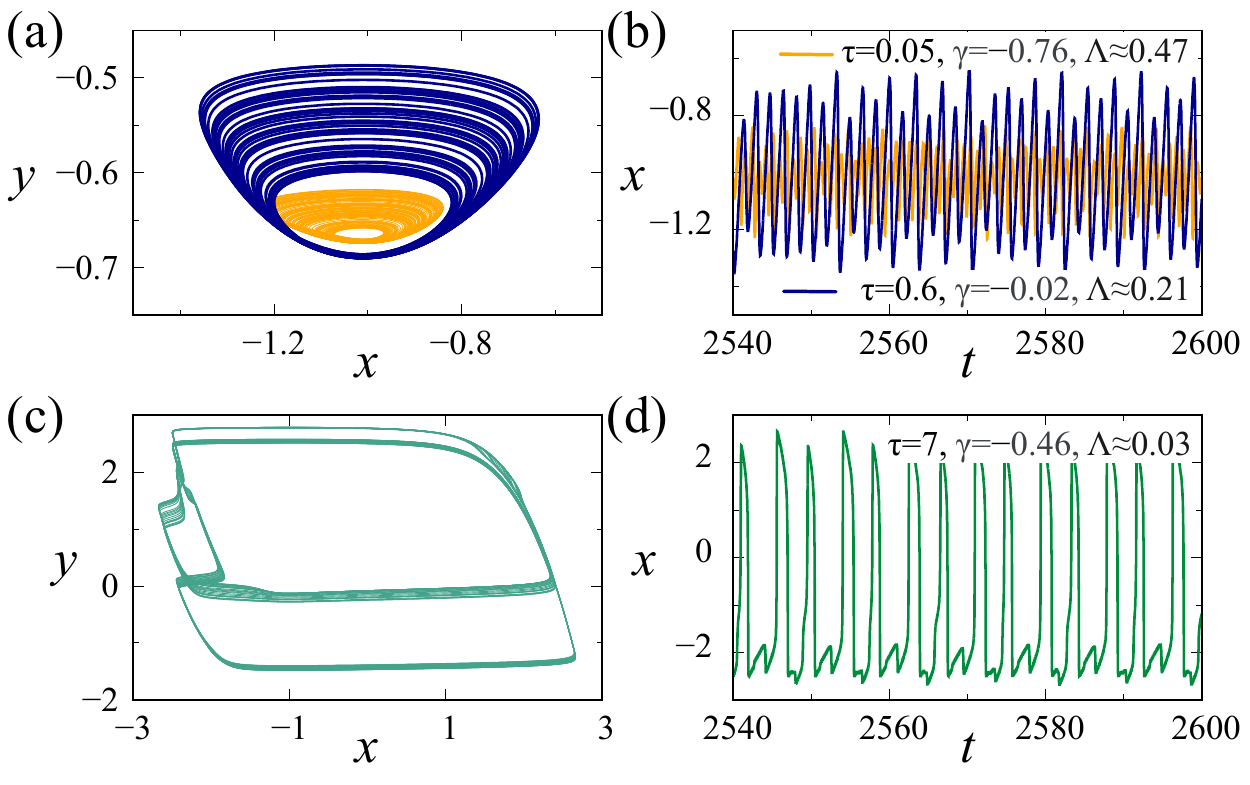}}
\caption{Three main types of chaotic regimes deminstrated by model (\ref{eq:system}) shown by projections of attractors on the phase plane $(x,y)$ in panels (a,c) with corresponding temporal dynamics (b,d). Legends contain the information about parameters and maximal Lyapunov exponent $\Lambda=\Lambda_1$. The rest parameters are $\varepsilon=0.05$, $a=1.01$.
% Phase portraits (a,c) and time series (b,d) of chaotic regimes in system (\ref{eq:system}) at $\varepsilon=0.05$, $a=1.01$, $\tau=0.05$, $\gamma=-0.76$ with $\Lambda_1\approx 0.47$ (orange curves in panels (a,b)); at $\tau=0.6$, $\gamma=-0.02$ with $\Lambda_1\approx 0.21$ (blue curves in panels (a,b)); and at $\tau=7$, $\gamma=-0.46$ with $\Lambda_1\approx 0.03$ (c,d).
}\label{fig:g_neg_PP}
\end{figure}

The exact parameter values for which chaotic regimes can be observed, do not change significantly for $a=1.01$ and $a=1.001$. The similar calculations for $a=1.001$ are given in \ref{sec:appendix:lyaps_gamma_negative}, Fig.~\ref{fig:g_neg_lyap_another_a}. In both $\gamma$ scales of Fig.~\ref{fig:g_neg_lyap}, there are very tiny areas with $\Lambda>0$ for $\tau<4$. The corresponding representative regimes are illustrated by projection of attracting sets on phase plane ($x,y$) and temporal dynamics in Fig.~\ref{fig:g_neg_PP} (a,b) for two values of delay time $\tau=0.05$ and $\tau=0.6$. In both cases, the system demonstrates chaotic oscillations of a small amplitude near the previously stable focus equilibrium point. Increasing the delay time $\tau$ one can observe a growth in the amplitude of chaotic oscillations (Fig.~\ref{fig:g_neg_PP}(c,d)). This is accompanied with enlarging the attractor in $(x,y)$ phase plane.

It is shown above that forward increase in $\tau$ value may transform the dynamics to the spiking regime (Fig.~\ref{fig:g_neg_realizations}). The phase portrait and temporal evolution in Fig.~\ref{fig:g_neg_PP}(c,d) show that system (\ref{eq:system}) may demonstrate chaotic spiking behaviour. Positive value of the maximal Lyapunov exponent $\Lambda\approx 0.04$ confirms this fact.

\begin{figure}[t]
\centering{\includegraphics[width=1\linewidth]{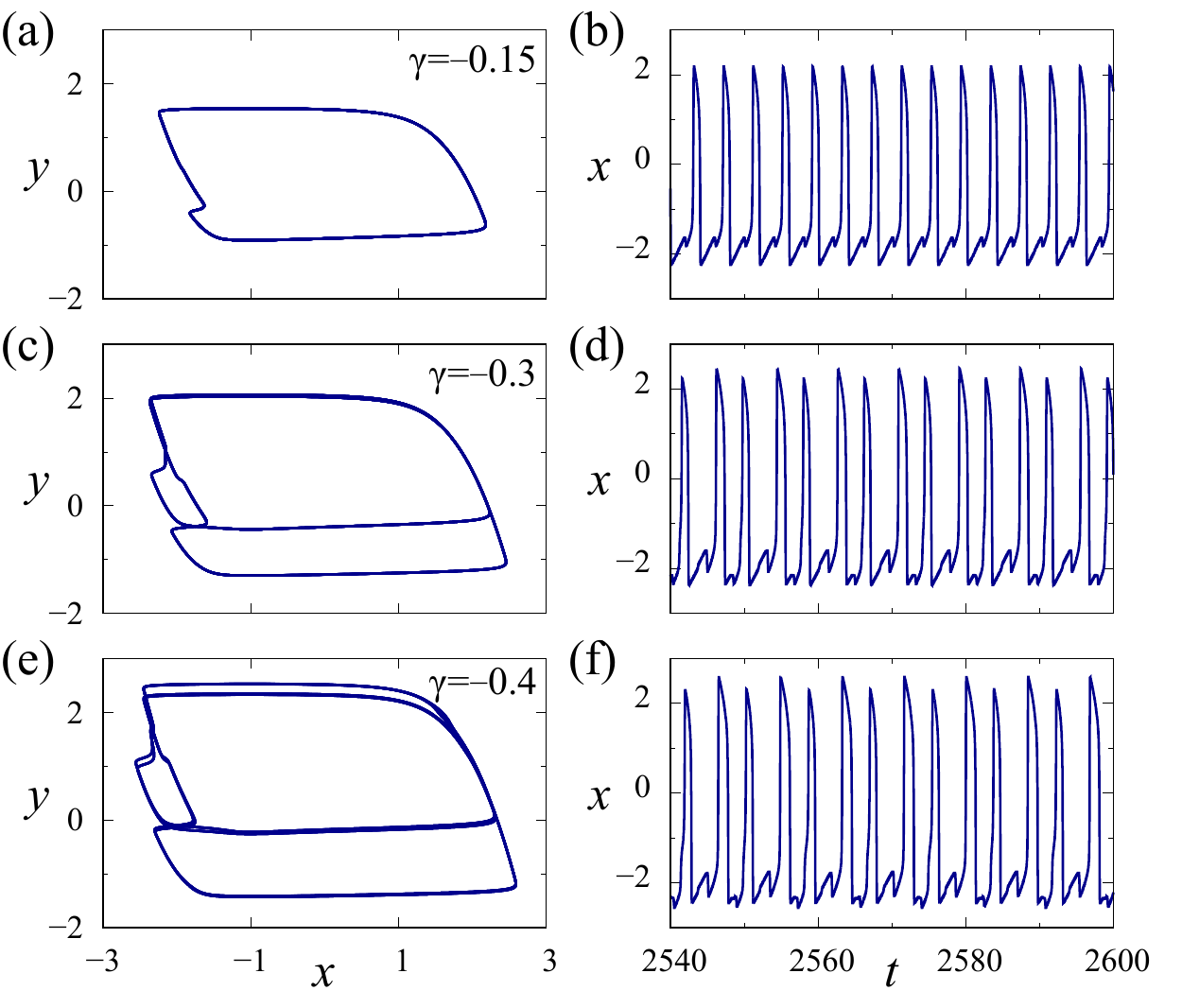}}
\caption{Transition to the chaotic behaviour through the cascade of period-doubling bifurcations in model (\ref{eq:system}) with negative feedback strength $\gamma$ illustrated by projections of attractors on the phase plane $(x,y)$ in panels (a,c,e) with corresponding temporal dynamics (b,d,f). This is accompanied with $\Lambda_1\approx 0.00$. Parameters: $\tau=7$, $\varepsilon=0.05$, $a=1.01$.
% Phase portraits (a,c,d) and time series (b,d,f) of chaotic regimes in system (\ref{eq:system}) at $\varepsilon=0.05$, $a=1.01$, $\tau=7$, $\gamma=-0.15$ with $\Lambda_1\approx 0.00$ (a,b); at $\gamma=-0.3$ with $\Lambda_1\approx 0.00$ (c,d); at $\gamma=-0.4$ with $\Lambda_1\approx 0.00$ (e,f).
}\label{fig:g_neg_PP_bifurcation}
\end{figure}

\begin{figure}[b]
\centering{\includegraphics[width=1\linewidth]{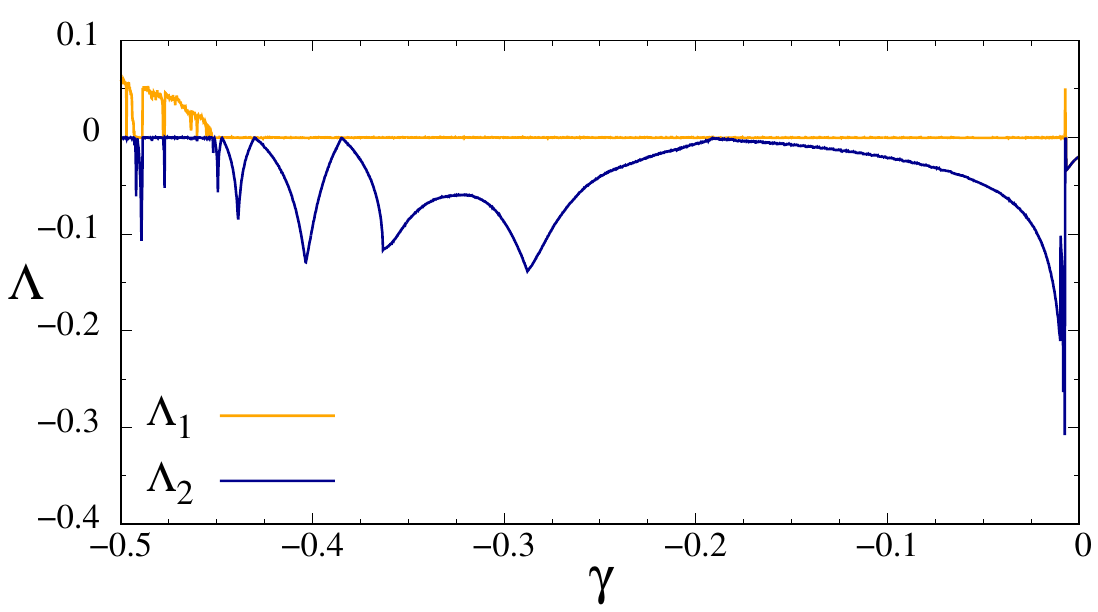}}
\caption{Two values of the Lyapunov spectrum: $\Lambda_1$ (orange) and $\Lambda_2$ (blue) depending on the negative $\gamma$ in model (\ref{eq:system}). Parameters: $\tau=7$, $\varepsilon=0.05$, $a=1.01$.}\label{fig:g_neg_lyaps_bifurcation}
\end{figure}

In order to examine the mechanism of the chaos appearance we now consider the behaviour of the system (\ref{eq:system}) during transition from regular periodic regime at $\tau=7$, $\gamma=-0.15$, $a=1.01$ to the largest region of existing chaotic regimes for $\tau\in(7;9)$ and $\gamma\in(-1;-0.4)$ shown in Fig.~\ref{fig:g_neg_lyap}(b). The projections of attracting sets on phase plane ($x,y$) and temporal dynamics obtained during this transition are shown in Fig.~\ref{fig:g_neg_PP_bifurcation} and demonstrate that the transition to the chaotic behaviour appears through the cascade of period-doubling bifurcations. The dependencies of the first and the second Lyapunov exponents on the parameter $\gamma$ are given in Fig.~\ref{fig:g_neg_lyaps_bifurcation}. The second Lyapunov exponent reaches zero value every time when the period of cycle is doubled. The Feigenbaum relations 
\begin{equation}\label{eq:Feigenbaum}
\delta_n = \dfrac{\gamma_{n}-\gamma_{n+1}}{\gamma_{n+1}-\gamma_{n+2}},
\end{equation}
calculated in Tab.~\ref{tab:Feigenbaum} are relatively close (with error about 7\% for $\delta_1$ and $\delta_3$) to value $\delta=4.669$ which correspond to $\lim_{n\to\infty}\delta_n$ in one-parameter maps. Finally, at $\gamma_{tr}=-0.45$ the system (\ref{eq:system}) comes to the chaotic regime, has been previously shown in Fig.~\ref{fig:g_neg_PP}(c,d).

\begin{table}[h]
\centering
\begin{tabular}{ | c | c | c | c | }
\hline
 period, $2^n$ & $\gamma_n$ & relations $\delta_n$ & comparing with $\delta$ \\ 
\hline
 $2^1=2$ & -0.1878 & 4.335 & 7.16\% \\  
\hline
 $2^2=4$ & -0.3846 & 2.64 & 43.47\% \\
\hline
 $2^3=8$ & -0.43 & 4.3 & 7.91\% \\
\hline
 $2^4=16$ & -0.4472 & -- & -- \\
\hline
 $2^5=32$ & -0.4512 & -- & -- \\
\hline
\end{tabular}
\caption{\label{tab:Feigenbaum}Feigenbaum relation values (\ref{eq:Feigenbaum}) characterizing transition to chaotic regime in model (\ref{eq:system}) with negative feedback strength.}
\end{table}

\subsection{Regular and chaotic waves at long delay}\label{sec:waves}
\begin{figure}[t] 
\centering{\includegraphics[width=0.9\linewidth]{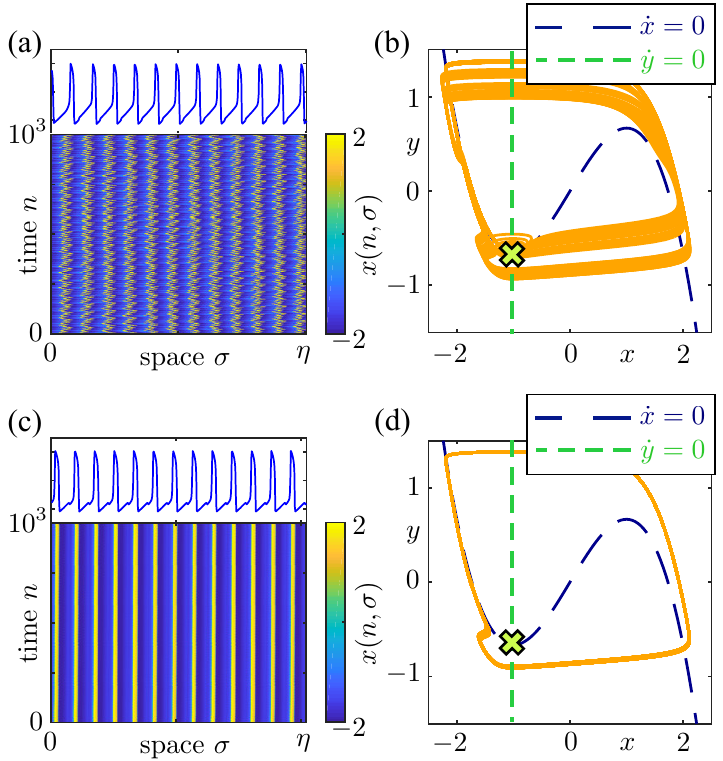}} 
\caption{Coexisting chaotic (a,b) at $\eta=50.3668$ and regular (c,d) self-oscillatory dynamics at $\eta=51.0155$ exhibited by model (\ref{eq:system}) with negative delay feedback strength. These regimes are illustrated by spatio-temporal diagrams (a,c) and by projections of corresponding attractors on ($x,y$) phase plane (b,d). Parameters: $\gamma=-0.1$, $\varepsilon=0.05$, $a=1.01$, $\tau=50$.
% Coexisting chaotic (panels (a) and (b)) and regular (panels (c) and (d)) self-oscillatory dynamics exhibited by model (\ref{eq:system}) at negative delay feedback strength. Panels (a) and (c) are the spatio-temporal diagrams while the corresponding phase portraits are depicted in panels (b) and (d). Parameters are: $\gamma=-0.1$, $\varepsilon=0.05$, $a=1.01$, $\tau=50$, $\eta=50.3668$ (panel (a)) and $\eta=51.0155$ (panel (c)).
}\label{fig:long_delay_negative} 
\end{figure}

For negative $\gamma$ and long time delay the represented in quasi-space dynamics of system (\ref{eq:system}) is  characterised by the excitation of wave-like patterns where the spatial impulses homogeneously fill the quasi space. Such structures do not involve the quiescent steady state regimes. This indicates that the delayed-feedback represents a factor for the steady state's instability at negative values $\gamma$. In contrast to the soliton structures obtained for positive $\gamma$, here one cannot  induce a certain number of impulses by varying the initial conditions. The coexistence of chaotic [Fig. \ref{fig:long_delay_negative} (a,b)] and regular [Fig. \ref{fig:long_delay_negative} (c,d)] self-oscillatory dynamics is achieved for big enough absolute values of the delay feedback strength. 

\section{Conclusion and discussion}
\label{sec:conclusion}
Delayed-feedback provides for the self-sustained oscillation excitation in the FitzHugh-Nagumo model. In our paper, we demonstrate that the delay-induced self-oscillatory dynamics can be regular and chaotic. In particular, at positive delayed-feedback strength one observes the bistability: the quiescent steady state regime coexists with regular self-oscillatory motions along the limit cycle. Despite the presented analysis of basins of attraction is reduced and involves the special kind of initial conditions $x_0=x(t\in[-\tau:0])=$const, it allows to establish that both the steady state and the limit cycles are characterised by finite basins of attraction.  Thus, they can be called 'attractors' in a full sense. 

For long time delay and positive delay feedback strength one can interpret the induced self-oscillations as dissipative solitons in the quasi-space. In such a case, the spatio-temporal dynamics involves both attractors and the delay-induced bistability is reflected in the ability to realize a certain number of solitons at certain times. 

For positive delayed-feedback strength the behaviour is more complicated: after the steady state loses its stability, both regular and chaotic self-oscillations can be achieved. For long time delay, regular and chaotic self-oscillations are manifested in the quasi-space as wave patterns. Surprisingly, increasing the absolute value of the negative delayed-feedback strength, one can realize the route to chaos through a cascade of period-doubling bifurcations such that the Feigenbaum constants registered in numerical experiments are close to the theoretical value $\delta=4.669$. The exhibition of the Feigenbaum scenario by the delayed-feedback FitzHugh-Nagumo model seems to be a non-trivial results and will be theoretically analysed in further studies. 

The obtained results are in a full correspondence with materials presented in Refs. \cite{ERN16,weicker2016,romeira2016} and other publications and complement a manifold of observed regimes by the chaotic dynamics.   

\section*{DATA AVAILABILITY}
The data that support the findings of this study are available from the corresponding author upon reasonable request.

\section*{Acknowledgements}
The main idea of the paper and the parts with probabilities, basins of attraction and linear stability analysis were prepared by N. Semenova, supported by The Council for grants of President of Russian Federation (project no. SP-749.2022.5). The calculations of Lyapunov exponents with finding the chaotic regimes were carried out by A. Bukh, supported provided by The Council for grants of President of Russian Federation (project no. SP-774.2022.5). Sects.~\ref{sec:solitons},~\ref{sec:waves} and the main work with the text was done by V. Semenov, supported by the Russian Science Foundation (project no. 22-72-00038). 

%% The Appendices part is started with the command \appendix;
%% appendix sections are then done as normal sections
\appendix

\section{Lyapunov exponents for positive time-delayed feedback strength $\gamma>0$}
\label{sec:appendix:lyaps_gamma_positive}
\begin{figure}[t] 
\centering{\includegraphics[width=1\linewidth]{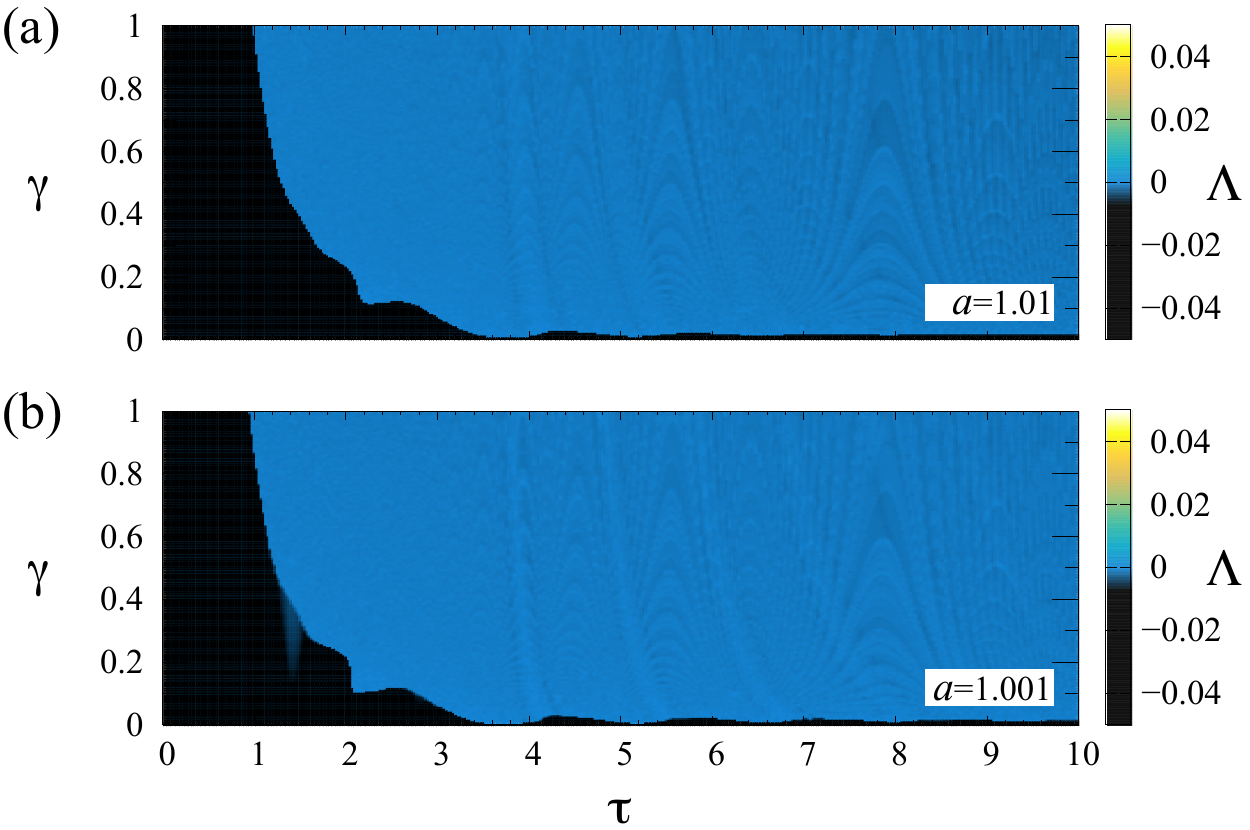}} 
\caption{Maximal Lyapunov exponent $\Lambda=\Lambda_1$ in ($\tau$,$\gamma$)-plane calculated for model (\ref{eq:system}) with positive feedback strength. Black area: stable equilibrium point; blue region: limit cycles. Parameters: $\varepsilon=0.05$, $a=1.01$ (a) and $a=1.001$ (b).
% Plane ($\tau$, $\gamma$) showing the values of the maximal Lyapunov exponent $\Lambda=\Lambda_1$ in the system (\ref{eq:system}) at $\varepsilon=0.05$ and $a=1.01$ (panel (a)) and $a=1.001$ (panel (b)), in case of positive feedback coupling. Black area corresponds to the stable equilibrium, yellow (light-grey) area confirms to limit cycles.
}\label{fig:g_pos_lyap_large_scale} 
\end{figure}

\section{Linear stability analysis for negative time-delayed feedback strength $\gamma<0$}
\label{sec:appendix:LSA_gamma_negative}

The FHN system in excitable regime ($a>1$) without time delay demonstrates the stable equilibrium point of type ``focus''. With decreasing parameter $a$ passing through $a=1$ this equilibrium point looses stability and becomes unstable. This can be illustrated by the roots of characteristic equation. The stable focus is accompanied by the complex roots $\lambda_{12}$ with negative real parts $\Re(\lambda_{12})<0$ in excitable regime ($a>1$), and with positive $\Re(\lambda_{12})>0$ in oscillatory regime ($0<a<1$). The bifurcation between these two regimes is accompanied with the case when the roots are purely imaginary with $\Re(\lambda_{12})=0$. This equilibrium type is called ``center'' \cite{Bashkirtseva2014}.

In the case of delayed system (\ref{eq:system}), the similar analysis can be applied. As it has been mentioned in the book \cite{LAK11}, in the general case the characteristic equation for the system with time-delayed feedback can be obtained as follows: 
\begin{equation}\label{eq:det_general} 
\mathrm{det}\left| \mathbf{J}+\mathbf{J_\tau}\cdot e^{-\lambda\tau}-\lambda\cdot \mathbf{I}\right|=0, 
\end{equation} 
where $\mathbf{J}$ is the Jacobian matrix for considered system without time-delayed terms, $\mathbf{J_\tau}$ is the Jacobian matrix for time-delayed terms, $\mathbf{I}$ is the unity matrix, $\lambda$ are the roots of characteristic equation, which type determines the type of equilibria. In the case of our system (\ref{eq:system}) these matrices are: 
\begin{equation}\label{eq:J}
\mathbf{J} = 
\begin{bmatrix}
 \frac{\partial f_x}{\partial x}  &  \frac{\partial f_x}{\partial y}  \\
 \frac{\partial f_y}{\partial x}  &  \frac{\partial f_y}{\partial y}  
\end{bmatrix}
=
\begin{bmatrix}
 (1-x^2_s-\gamma)/\varepsilon & -1/\varepsilon \\
 1 & 0 
\end{bmatrix},
\end{equation}
where $f_x$ and $f_y$ are the right and left parts of Eq.~(\ref{eq:system}), $x_s$ is the $x$-value of the equilibrium state equal to $x_s=-a$. The Jacobian matrix for delayed part is:
\begin{equation}\label{eq:Jtau}
\mathbf{J_\tau} = 
\begin{bmatrix}
 \frac{\partial f_x}{\partial x_\tau}  &  \frac{\partial f_x}{\partial y_\tau}  \\
 \frac{\partial f_y}{\partial x_\tau}  &  \frac{\partial f_y}{\partial y_\tau}  
\end{bmatrix}
=
\begin{bmatrix}
 \gamma/\varepsilon & 0 \\
 0 & 0 
\end{bmatrix}.
\end{equation}
Let us denote $b=a^2-1$, which is positive when $a>1$. Then the characteristic equation is:
\begin{equation}\label{eq:character_eq_general_matrix}
\mathrm{det}
\begin{bmatrix}
 -\frac{1}{\varepsilon}(b+\gamma)-\lambda+\frac{\gamma}{\varepsilon} e^{-\lambda\tau} & -\frac{1}{\varepsilon} \\
 1 & -\lambda 
\end{bmatrix}=0
\end{equation}
\begin{equation}\label{eq:character_eq_general_line}
\varepsilon\lambda^2+ \lambda(b+\gamma)-\gamma\lambda e^{-\lambda\tau} +1 =0.
\end{equation}
The purely imaginary $\lambda$ corresponds to the equilibrium of type ``center''. Therefore, we set $\lambda=i\omega$, $\omega\in\mathbb{R}$, $\omega\neq 0$. Then $e^{-\lambda\tau}$ can be rewritten as $(\cos\omega\tau-i\cdot\sin\omega\tau)$ and (\ref{eq:character_eq_general_line}) transforms to
\begin{equation}\nonumber
i\omega(b+\gamma)-\varepsilon\omega^2-i\omega\gamma\cos\omega\tau-\omega\gamma\sin\omega\tau+1=0
\end{equation}
Dividing this equation into a system of two equations for the real and imaginary parts, we obtain
\begin{equation}\label{eq:sin_cos_system}
\begin{cases}
\cos\omega\tau = \frac{b+\gamma}{\gamma}\\
\sin\omega\tau = \frac{1-\varepsilon\omega^2}{\omega\gamma}
\end{cases}
\end{equation}
The first equation can be rewritten in the form $\gamma=-b/(1-\cos\omega\tau)$, indicating that $\gamma<0$ when $a>1$. Therefore, next we will consider only the case of negative $\gamma$.

To find a solution $\omega$ to the system (\ref{eq:sin_cos_system}), we use the rule $\sin^2\omega\tau+\cos^2\omega\tau=1$:
\begin{equation}
\begin{array}{c}
\frac{(b+\gamma)^2}{\gamma^2} + \frac{(1-\varepsilon\omega^2)^2}{\omega^2\gamma^2} = 1, \\
\varepsilon^2\omega^4+\omega^2\big(b(b+2\gamma)-2\varepsilon\big)+1=0.
\end{array}
\end{equation}
Denote $\xi=b(b+2\gamma)-2\varepsilon$. The parameters, which are used in this paper, can lead only to $\xi<0$. Then the equation trainsforms to 
\begin{equation}\label{eq:xi_equation}
\varepsilon^2\omega^4 + \xi\omega^2+1=0
\end{equation}
with the discriminant $D=\xi^2-4\varepsilon^2$. The solution exists only if $D\ge 0$. And this leads to additional conditions for $\xi$: $\xi\ge 2\varepsilon$ or $\xi\le -2\varepsilon$. Combining this with $\xi<0$, we get the final condition:
\begin{equation}\label{eq:xi_condition}
\xi\le -2\varepsilon.
\end{equation}
Equation (\ref{eq:xi_equation}) produces four solutions $\omega_1,\ \omega_3>0$ and $\omega_2,\ \omega_4>0$:
\begin{equation}\label{eq:omega_1234}
\left[
\begin{array}{l}
\omega^2_{1,2} = \frac{-\xi+\sqrt{D}}{2\varepsilon^2} \\
\omega^2_{3,4} = \frac{-\xi-\sqrt{D}}{2\varepsilon^2} 
\end{array}
\right.
\end{equation}
All four solutions exist when condition (\ref{eq:xi_condition}) holds. However, in what follows we will need to compare $\omega^2$ with the value $1/\varepsilon$. 

~\\
\textbf{1)} $\omega^2\ge 1/\varepsilon$: \\
Let us find the condition for existence $\omega_{1,2}$. The inequality $\frac{-\xi+\sqrt{D}}{2\varepsilon^2}\ge \frac{1}{\varepsilon}$ simply transforms to $\sqrt{D}\ge 2\varepsilon+\xi$. This leads to the system of inequalities:
\begin{equation}\label{eq:w12_condition1}
\left[
\begin{array}{l}
\begin{cases}
2\varepsilon+\xi\ge 0 \\
D\ge (2\varepsilon+\xi)^2
\end{cases}
\\
\begin{cases}
2\varepsilon+\xi< 0 \\
D\ge 0
\end{cases}
\end{array}
\right.
\Leftrightarrow
\left[
\begin{array}{l}
\xi=-2\varepsilon \\
\xi<-2\varepsilon
\end{array}
\right.
\Leftrightarrow
\xi\le-2\varepsilon.
\end{equation}
Similarly, one can find a condition for the existence of $\omega_{3,4}$ from $\frac{-\xi-\sqrt{D}}{2\varepsilon^2}\ge \frac{1}{\varepsilon}$:
\begin{equation}\label{eq:w34_condition1}
\begin{cases}
D\le (\xi+2\varepsilon)^2 \\
D\ge 0 \\
-\xi-2\varepsilon\ge 0
\end{cases}
\Leftrightarrow
\begin{cases}
\xi\ge -2\varepsilon \\
\xi\le -2\varepsilon
\end{cases}
\Leftrightarrow
\xi=-2\varepsilon.
\end{equation}
The case $\xi=-2\varepsilon$ corresponds to $D=0$ and $\omega_{1,3}=\sqrt{-\xi/(2\varepsilon^2)}$, $\omega_{2,4}=-\sqrt{-\xi/(2\varepsilon^2)}$. Therefore, the condition (\ref{eq:w34_condition1}) has already been taken into account in condition (\ref{eq:w12_condition1}). 

Substituting all the renaming into (\ref{eq:w12_condition1}), we get $\gamma\le -b/2$.
Thus the condition $\omega^2\ge1/\varepsilon$ means that $\gamma\le -b/2$ and then only $\omega_{1,2}$ exists.
~\\
~\\
\textbf{2)} $\omega^2\le 1/\varepsilon$: \\
Similarly to the previous calculations, we get that condition $\omega^2\le 1/\varepsilon$ means that $\gamma\le -b/2$ and then only $\omega_{3,4}$ exists. This is illustrated by the next system of inequalities:
\begin{equation}\label{eq:omega_epsi}
\begin{array}{c}
\omega^2\ge 1/\varepsilon
\Leftrightarrow
\begin{cases}
\gamma\le -b/2\\
\left[
\begin{array}{l}
\omega=\omega_1,\ \omega> 0\\
\omega=\omega_2,\ \omega< 0
\end{array}
\right.
\end{cases} \\
\omega^2\le 1/\varepsilon
\Leftrightarrow
\begin{cases}
\gamma\le -b/2\\
\left[
\begin{array}{l}
\omega=\omega_3,\ \omega> 0\\
\omega=\omega_4,\ \omega< 0
\end{array}
\right.
\end{cases}
\end{array}
\end{equation}
~\\
Let us return to the system of equations (\ref{eq:sin_cos_system}) and consider the solutions depending on which quarter of the unity circle the angle $\omega\tau$ belongs to.

\textbf{I)} The first quarter: $0+2\pi n\le\omega\tau\le \pi/2+2\pi n$, $n\in  \mathbb{Z}$.
\begin{equation}
\begin{cases}
  \cos\omega\tau\ge 0 \\
  \sin\omega\tau\ge 0
\end{cases}
\Leftrightarrow
\begin{cases}
  \gamma \le -b \\
  \left[
  \begin{array}{l}
    \begin{cases}
      \omega<0 \\
      \omega^2\le 1/\varepsilon
    \end{cases}\\
    \begin{cases}
      \omega>0 \\
      \omega^2\ge 1/\varepsilon
    \end{cases}
  \end{array}
  \right.
\end{cases}
\end{equation}
Including (\ref{eq:omega_epsi}), we get
\begin{equation}\label{eq:1quarter}
\begin{cases}
  \gamma\le -b \\
  \left[
  \begin{array}{l}
    \omega=\omega_1 \\
    \omega=\omega_4
  \end{array}
  \right.
\end{cases}
\end{equation}

\textbf{II)} The second quarter: $\pi/2+2\pi n\le\omega\tau\le \pi+2\pi n$, $n\in  \mathbb{Z}$.
\begin{equation}\label{eq:2quarter}
\begin{cases}
  \cos\omega\tau\le 0 \\
  \sin\omega\tau\ge 0
\end{cases}
\Leftrightarrow
\begin{cases}
  -b\le\gamma\le -b/2 \\
  \left[
  \begin{array}{l}
    \omega=\omega_1 \\
    \omega=\omega_4
  \end{array}
  \right.
\end{cases}
\end{equation}

\textbf{III)} The third quarter: $\pi+2\pi n\le\omega\tau\le 3\pi/2+2\pi n$, $n\in  \mathbb{Z}$.
\begin{equation}\label{eq:3quarter}
\begin{cases}
  \cos\omega\tau\le 0 \\
  \sin\omega\tau\le 0
\end{cases}
\Leftrightarrow
\begin{cases}
  -b\le\gamma\le -b/2 \\
  \left[
  \begin{array}{l}
    \omega=\omega_2 \\
    \omega=\omega_3
  \end{array}
  \right.
\end{cases}
\end{equation}

\textbf{IV)} The fourth quarter: $3\pi/2+2\pi n\le\omega\tau\le 2\pi(n+1)$, $n\in  \mathbb{Z}$.
\begin{equation}\label{eq:4quarter}
\begin{cases}
  \cos\omega\tau\ge 0 \\
  \sin\omega\tau\le 0
\end{cases}
\Leftrightarrow
\begin{cases}
  \gamma\le -b \\
  \left[
  \begin{array}{l}
    \omega=\omega_2 \\
    \omega=\omega_3
  \end{array}
  \right.
\end{cases}
\end{equation}

One can simply get the equation for $\gamma$ from (\ref{eq:sin_cos_system}):
\begin{equation}\label{eq:gamma}
\gamma = -\frac{b}{1-\cos\omega\tau}.
\end{equation}
The solution $\omega$ depends on the parameter $\gamma$, and the bifurcation line can be therefore plotted in the parameter plane ($\gamma,\tau$), if Eq.~(\ref{eq:gamma}) is solved graphically. The function $\cos$ is even and it is enough to consider only $\omega\tau$-quarters with existing $\omega_1$ and $\omega_4$.
Figure~\ref{fig:appendix_analytics} contains the graphical solution of the next systems of equations and inequalities:
\begin{figure}[t] 
\centering{\includegraphics[width=1.0\linewidth]{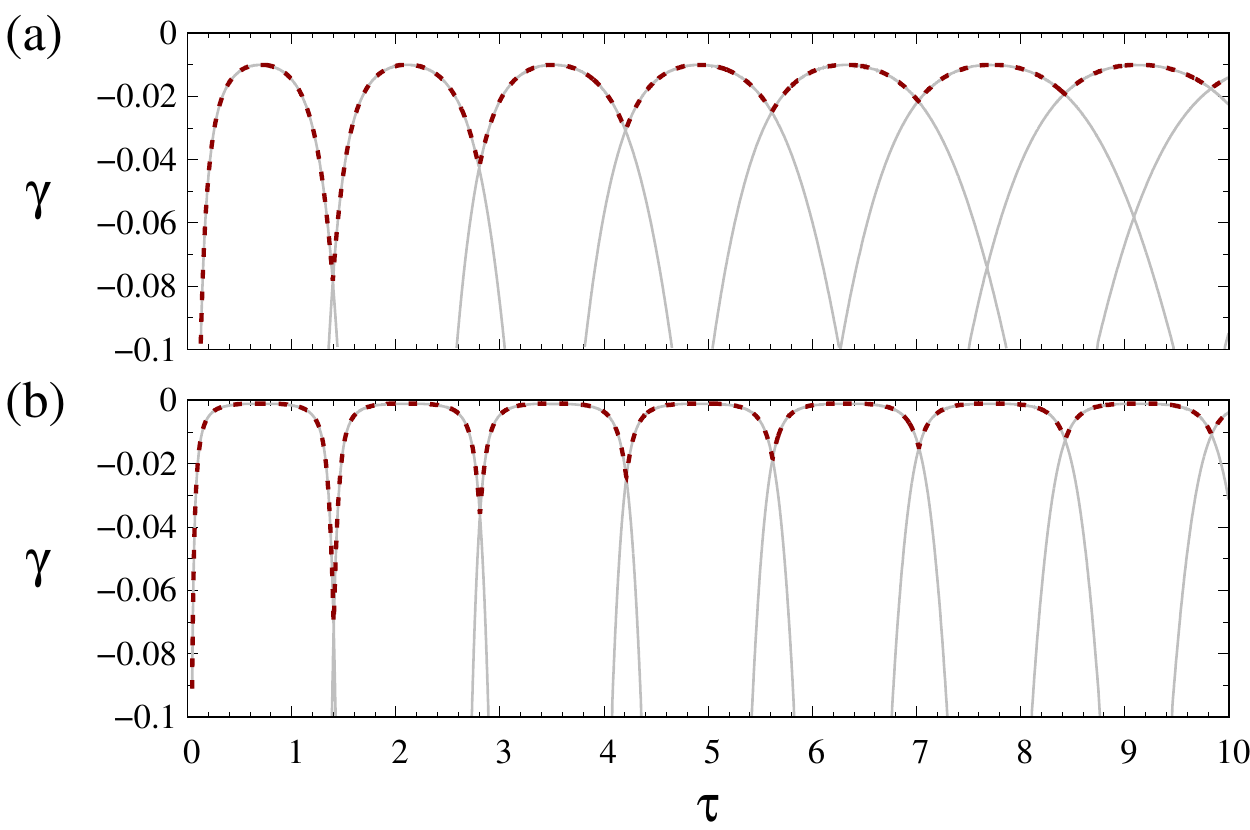}} 
\caption{The solution of system (\ref{eq:gamma}) for equilibrium state of type ``circle'' on ($\tau$, $\gamma$)-plane. Parameters: $\varepsilon=0.05$, $a=1.01$ (a) and $a=1.001$ (b).
% Plane ($\tau$, $\gamma$) showing the solution of the system (\ref{eq:gamma}), when the FitzHugh--Nagumo demonstrates the equilibrium state of type ``circle''. Parameters: $\varepsilon=0.05$, $a=1.01$ (a) and $a=1.001$ (b).
}\label{fig:appendix_analytics}
\end{figure} 
\begin{equation}\nonumber
\begin{cases}
\gamma\le -b \\
\left[
\begin{array}{l}
  \gamma = -\frac{b}{1-\cos\omega_1\tau}, \ \ \ \mathrm{if}\ \ \omega_1\tau\in \mathrm{\ I^{st}\ quarter} \\
  \gamma = -\frac{b}{1-\cos\omega_4\tau}, \ \ \ \mathrm{if}\ \ \omega_4\tau\in \mathrm{\ I^{st}\ quarter}
\end{array}
\right.
\end{cases}
\end{equation}
and
\begin{equation}\nonumber
\begin{cases}
-b\le\gamma\le -b/2 \\
\left[
\begin{array}{l}
  \gamma = -\frac{b}{1-\cos\omega_1\tau}, \ \ \ \mathrm{if}\ \ \omega_1\tau\in \mathrm{\ II^{nd}\ quarter} \\
  \gamma = -\frac{b}{1-\cos\omega_4\tau}, \ \ \ \mathrm{if}\ \ \omega_4\tau\in \mathrm{\ II^{nd}\ quarter}
\end{array}
\right.
\end{cases}
\end{equation}
where $b=a^2-1$, $\xi=b(b+2\gamma)-2\varepsilon$,\\ $\omega_1=\sqrt{\frac{-\xi+\sqrt{\xi^2-4\varepsilon^2}}{2\varepsilon^2}}$,\\ $\omega_4=-\sqrt{\frac{-\xi-\sqrt{\xi^2-4\varepsilon^2}}{2\varepsilon^2}}$

The dashed line in Fig.~\ref{fig:appendix_analytics} separates the area upper the bifurcation line with stable focus and the lower area with unstable focus. This line purely coinsides with the results of numerical simulation (Fig.~\ref{fig:g_neg_prob}).

\section{Lyapunov exponents for positive time-delayed feedback strength $\gamma<0$}
\label{sec:appendix:lyaps_gamma_negative}
\begin{figure}[t] 
\centering{\includegraphics[width=0.98\linewidth]{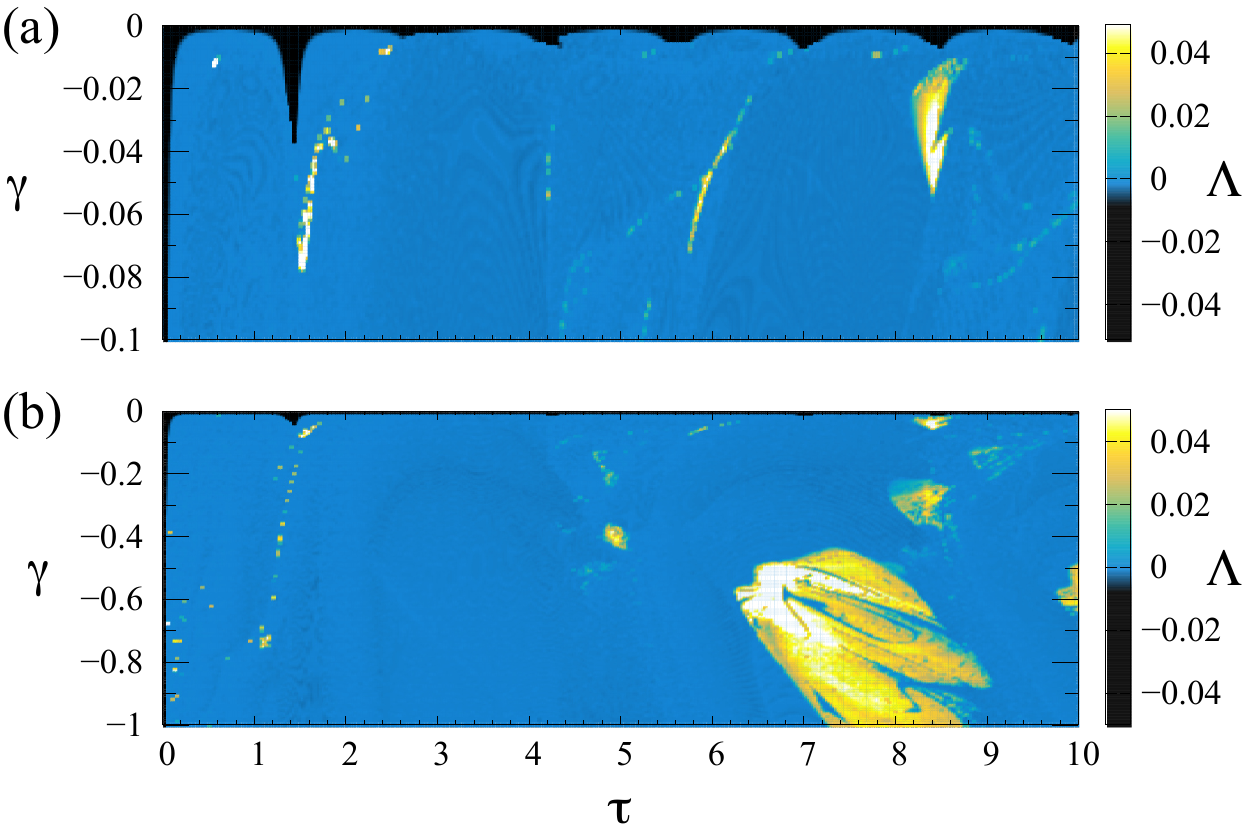}} 
\caption{Maximal Lyapunov exponent $\Lambda=\Lambda_1$ in ($\tau$,$\gamma$)-plane calculated for model (\ref{eq:system}) with negative feedback strength for two different ranges of $\gamma$ values. Black area: stable equilibrium point; blue region: limit cycles; yellow or white areas: chaotic regimes. Parameters: $\varepsilon=0.05$, $a=1.001$.
% Plane ($\tau$, $\gamma$) showing the values of the maximal Lyapunov exponent $\Lambda=\Lambda_1$ in system (\ref{eq:system}) at $\varepsilon=0.05$ and $a=1.001$, in case of negative feedback coupling. Black area corresponds to the stable equilibrium, blue area confirms to limit cycles, and yellow area contain chaotic regimes.
}\label{fig:g_neg_lyap_another_a} 
\end{figure}

%% If you have bibdatabase file and want bibtex to generate the
%% bibitems, please use
%%
%\bibliographystyle{elsarticle-num} 
%\bibliography{ref}

\end{document}